\documentclass[nohyper,notoc]{JHEP3}
\usepackage{amsmath,amssymb,amsfonts,epsfig}

\def\be{\begin{equation}}
\def\ee{\end{equation}}
\def\bea{\begin{eqnarray}}
\def\eea{\end{eqnarray}}
\def\bse{\begin{subequations}}
\def\ese{\end{subequations}}

\newcommand\bra[1]{{\langle {#1}|}}
\newcommand\ket[1]{{|{#1}\rangle}}

\def\cN{\mathcal{N}}
\def\pa{\partial}
\def\Im{{\rm Im}}
\newcommand{\I}{{\mathrm i}}
\newcommand{\ihalf}{\ensuremath{{\frac{\I}{2}}}}
\newcommand{\dwrt}[1]{\frac{\partial}{\partial#1}}

\newcommand{\half}{\ensuremath{{\frac{1}{2}}}}
\newcommand{\IR}{\mathbb{R}}
\newcommand{\IZ}{\mathbb{Z}}
\newcommand{\hp}{{\wh{p}}}
\newcommand{\hq}{{\wh{q}}}
\newcommand{\wh}[1]{\widehat{#1}}

\title{Topological wave functions and the 4D-5D lift}

\author{Peng Gao\\
Department of Physics, University of Toronto,\\ 60 St. George st, Toronto, ON M5S 1A7, Canada\\
\\
{\tt E-mail: gaopeng@physics.utoronto.ca}
}

\author{Boris Pioline\\
Laboratoire de Physique Th\'eorique et Hautes
Energies\footnote{Unit\'e mixte de recherche du CNRS UMR 7589},\\
Universit\'e Pierre et Marie Curie - Paris 6,
4 place Jussieu, F-75252 Paris cedex 05 \\

Laboratoire de Physique Th\'eorique de l'Ecole Normale
Sup\'erieure\footnote{Unit\'e mixte
de recherche du CNRS UMR 8549},\\
24 rue Lhomond, F-75231 Paris cedex 05\\
\\
{\tt E-mail: pioline@lpthe.jussieu.fr}
}

\abstract{We revisit the holomorphic anomaly
equations satisfied by the topological string amplitude from the 
perspective of the 4D-5D lift, in the context of ``magic'' $\cN=2$
supergravity theories. In particular, we interpret the 
Gopakumar-Vafa relation between 5D black hole degeneracies and the 
topological string amplitude as the result of a canonical transformation
from 4D to 5D charges. Moreover we use the known Bekenstein-Hawking entropy
of 5D black holes to constrain the asymptotic behavior of the 
topological wave function at finite topological coupling but large K\"ahler
classes. In the process, some subtleties in the relation between 5D black hole 
degeneracies and the topological string amplitude are uncovered, 
but not resolved. Finally we extend these considerations 
to the putative one-parameter 
generalization of the topological string amplitude, and identify the
canonical transformation as a Weyl reflection inside the 3D duality group.}

\preprint{LPTENS-08/14\\ arXiv:0803.0562}

\begin{document}

\section{Introduction}\label{s1}
BPS black holes in $\cN=2$ supergravity theories have attracted  
revived attention recently, with the discovery of deep connections 
between topological strings and the 
entropy of four-dimensional \cite{Ooguri:2004zv} and
five-dimensional \cite{Gopakumar:1998ii,Katz:1999xq} 
black holes, as well as a direct relation between 4D black holes 
and 5D black holes and black rings, often known as the 4D-5D 
lift \cite{Bena:2004tk,Gaiotto:2005gf,Gaiotto:2005xt,Bena:2005ni}.
These advances have also led to much progress in our understanding 
of the topological string amplitude, in particular with regard
to its wave function character~\cite{Witten:1993ed,Verlinde:2004ck,
Gerasimov:2004yx,Aganagic:2006wq,Gunaydin:2006bz,Schwarz:2006br,Huang:2006hq}.

In this note, we revisit the holomorphic anomaly
equations satisfied by the topological string amplitude from the 
perspective of the 4D-5D lift. 
Our analysis is based on the construction in~\cite{Gunaydin:2006bz},
where the standard topological amplitude~\cite{Bershadsky:1993ta} (BCOV) was
recast into a purely holomorphic wave function satisfying 
a generalized heat equation, as first suggested in
\cite{Witten:1993ed} (see \cite{Schwarz:2006br} for a closely related 
construction). In the same work, 
the algebraic nature of the topological amplitude was elucidated 
in the context of so-called ``magic'' ${\cal N}=2$ supergravity 
theories~\cite{Gunaydin:1983rk}, characterized by the fact that their
moduli space is a symmetric space. Some of these models are known to be
consistent quantum $\cN=2$ theories \cite{Ferrara:1995yx},
while others arise as truncations of theories with higher 
supersymmetry (see \cite{Dolivet:2007sz,Bianchi:2007va}
for recent progress on this issue).

The outline of this note is as follows. In Section 2, we review the 
relevant results from \cite{Gunaydin:2006bz} pertaining to ``magic''
supergravities. In Section 3.1, we observe that the relation between
the charges of 4D and 5D black holes related by the 4D-5D lift is a 
canonical transformation. This motivates us to introduce a 
new ``5D'' polarization for the topological wave function, 
$\Psi_{5D}(Q_i,J)$, related to the standard ``real'' polarization 
$\Psi_{\IR}(p^I)$ by an appropriate Bogoliubov transformation. 
In Section 3.2, we show that this relation is an instance of
the Gopakumar-Vafa connection between 5D black holes and topological 
strings \cite{Gopakumar:1998ii,Katz:1999xq},
provided we identify $\Psi_{5D}(Q_i,J)$ with the degeneracies of 5D 
BPS black holes. In Section 3.3, by exploiting the known 
Bekenstein-Hawking-Wald entropy of 5D black holes, we constrain the 
asymptotic behavior of the topological string amplitude at finite coupling 
but for large K\"ahler classes.
In Section 4, we extend this 5D polarization 
to the putative ``generalized topological amplitude'' introduced in 
\cite{Gunaydin:2006bz}, which remains to be constructed, and identify the
canonical transformation as a particular Weyl reflection inside 
the 3D duality group. 
The details of two computations are relegated in Appendices A and B.

\section{Magic supergravities and topological wave functions}
In this section, we briefly review the main results in~\cite{Gunaydin:2006bz}
on the algebraic nature of the topological string amplitude in 
 ``magic'' $\cN=2, D=4$ supergravity theories. Some useful 
background can be found in \cite{Gunaydin:1983rk,Pioline:2006ni}.

In these models, the vector multiplet moduli space is a Hermitian 
symmetric tube domain  ${\cal M}=G/K$ (a very special case of a 
special K\"ahler manifold),
$G={\rm Conf}(J)$ is the ``conformal group'' associated to a Jordan algebra 
$J$ of degree three, 
$K$ is the maximal compact subgroup of $G$, a compact real form of
the ``reduced structure group'' 
${\rm Str}_0(J)$, and the role of the phase 
space $H^{\rm even}(X,\mathbb{R})$ in type IIA 
compactifications on a Calabi-Yau three-fold $X$
(or $H^3(X,\mathbb{R})$ in type IIB compactifications)
is played by the ``Freudenthal triple'' associated to $J$, namely
the real vector space 
\be
\label{RJJR}
V=\mathbb{R}\oplus J\oplus J\oplus\mathbb{R} 
\ee 
equipped with the symplectic form
\be
\label{ompq}
\omega =
dp^0 \wedge dq_0 + dp^i \wedge dq_i \equiv dp^I \wedge dq_I
\ee
where $(p^0,p^i,q_i,q_0)$ are the coordinates along the 
respective summands in \eqref{RJJR}. $V$ admits a linear action of $G$ 
which preserves the symplectic form $\omega$, and leaves the
quartic polynomial 
\be
\label{i4pq}
I_4(p^I,q_I) = 4 p^0 N(q_i) -4 q_0 N(p^i) + 4 \pa_{p^i}N(p^j)
\pa_{q_i}N(q_j) - (p^0 q_0 + p^i q_i)^2
\ee 
invariant. The symplectic space $V$
may be quantized by replacing $(p^I,q_I)$ by operators 
$\hat{p}^I=p^I,\hat{q}_I=\I\hbar\, \pa/\pa p^I$ acting
on the Hilbert space $\mathcal{H}$ of $L^2$ functions of $n_v+1$ 
variables $p^I$, generating the 
Heisenberg group $H$ with center $Z=-\I\hbar$,
\be 
[\hat{p}^I,\hat{q}_J]=Z\,\delta^I_J \ .
\ee 
The linear action of $G$ on $V$ leads to a unitary action 
of $G$ on $\mathcal{H}$ by generators in the universal enveloping 
algebra of $H$\footnote{For later convenience, we have flipped 
the sign of $\hat q_I$ and $Z$ with respect to \cite{Gunaydin:2006bz},
and reinstored a general value for $\hbar$.}
\bse
\label{swrep}
\begin{gather}
S^i \mapsto - \ihalf \hbar^2\, C^{ijk} \frac{\pa^2}{\pa p^j \pa p^k} 
- \hbar\, p^i \dwrt{p^0}, 
\qquad T_i \mapsto \ihalf C_{ijk} p^j p^k - \hbar\,p^0 
\dwrt{p^i}, \label{swrep-3} \\
R^j_i \mapsto -\delta^j_i\, \hbar\, p^0 \dwrt{p^0} + \hbar\, p^j \dwrt{p^i} 
- \half C_{ikl}\, C^{jnl}\,\hbar 
\left(p^k \dwrt{p^n} + \dwrt{p^n} p^k\right)\ ,
\label{swrep-4}\\
D\equiv \frac{3}{n_v} R^i_i 
\mapsto -3 \hbar\, p^0 \dwrt{p^0} - \hbar\, p^i \dwrt{p^i} 
- \half \hbar\, (n_v+3). \label{swrep-5}
\end{gather}
\ese
Here, $C_{ijk}$ is the cubic norm form of $J$, related to the prepotential
$F_0$ describing the vector multiplet moduli space $\mathcal{M}$ via
\be
F_0 = \frac{1}{6} \frac{C_{ijk} X^i X^j X^k}{X^0} \equiv \frac{N(X^i)}{X^0}\ ,
\ee
and $C^{ijk}$ is 
the ``adjoint norm form'', satisfying the ``adjoint identity''
\be
Q_i = \frac12 C_{ijk} Q^j\, Q^k \quad\Leftrightarrow\quad
Q^i =\frac{1}{\sqrt{N(Q_i)}} \,\frac12 C^{ijk} Q_j\, Q_k\ ,\quad
N(Q_i)\equiv \frac16 C^{ijk}Q_i Q_j Q_k\ .
\ee
Thus,  the Hilbert space $\mathcal{H}$ furnishes a unitary representation
of the ``Fourier-Jacobi group'' $\tilde G = G \ltimes H$, known
as the Schr\"odinger-Weil representation.

Moreover, in \cite{Gunaydin:2006bz} (without any assumption of ``magicness''), 
a sequence of transformations was constructed which takes the topological 
partition function $\Psi_{\rm BCOV}(t^i,\bar t^{\bar i};x^i,\lambda)$
from  \cite{Bershadsky:1993ta}, 
subject to two sets of holomorphic anomaly equations, into a purely holomorphic
wave function $\Psi_{\rm hol}(t^i; y_i,w)$ satisfying a single heat 
equation\footnote{This $\Psi_{\rm hol;GNP}(t^i;y_i,w)$ is related to the 
holomorphic wave function $\Psi_{\rm hol,ST}(t^i; \lambda,\epsilon^0,
\epsilon^i)$ introduced in \cite{Schwarz:2006br} by setting $\lambda=1$ 
using homogeneity, and Fourier transforming $(\epsilon^0,\epsilon^i)$ 
into $(w,y_i)$. $\Psi_{\rm hol,ST}$ arises as the holomorphic limit of 
the BCOV topological amplitude $\Psi_{\rm BCOV}(t^i,\bar t^{\bar i}
\to\infty,x^i,\lambda)$, whereas $\Psi_{\rm hol;GNP}$ may be obtained
directly from $\Psi_{\rm BCOV}$ without any limiting or integration
procedure.}
\be
\label{heat}
\left[ \frac{\pa}{\pa t^i} -\frac{\I}{2} C_{ijk} 
\frac{\pa^2}{\pa y_j \pa y_k} + y_i \frac{\pa}{\pa w}
\right]\, \Psi_{\rm hol}(t^i;y_i,w) = 0\ .
\ee
In magic cases, it was further shown that this 
holomorphic wave function can be viewed as a matrix element
\be
\Psi_{\rm hol}(t^i; y_i,w) = \langle \Psi | 
\exp\left( y_i \hat{p}^i + (w-t^i y_i) \hat{p}^0 \right)
\, \exp(t^i T_i) \, | \Omega_0 \rangle
\ee
where $|\Omega\rangle_0$ is the ``vacuum'' of the  Schr\"odinger-Weil 
representation, annihilated by $\hq_I$, $S_i$ and 
the traceless part of $R^i_j$, and with charges
\be
D\,|\Omega\rangle_0 = -\frac12 (n_v+3)|\Omega\rangle_0\ ,\quad
Z\,|\Omega\rangle_0 = -\I |\Omega\rangle_0\ .
\ee
The heat equation \eqref{heat} (and ultimately the 
holomorphic anomaly equations of \cite{Bershadsky:1993ta}) 
can then be shown to follow from the 
operator identity in the Schr${\rm \ddot o}$dinger-Weil
representation of $\tilde G$,
\be
\label{gnp} 
Z \,T_i=
\hat p^0\hat q_i +
\half \, C_{ijk}\,\hat p^j\hat p^k \ ,
\ee 
It is also useful to introduce the operator 
\be
\label{gnpJ}
2 \hat J = \frac23 Z \,\hp^i\,T_i + \hat p^0 \,\left(
\hat p^0 \hat q_0 + \frac13 \, \hat p^i \hat q_i \right)
= 
\hat p^0 ( \hat  p^0 \hat  q_0 + \hat  p^i \hat  q_i) 
+ \frac13\, C_{ijk}\,\hat p^i\,\hat p^j\,\hat p^k\ ,
\ee
whose significance will become apparent shortly.

\section{Topological wave functions and black hole entropy}\label{s3}

Our starting point is the observation that the right-hand sides 
of \eqref{gnp} and \eqref{gnpJ} formally give the electric charges $Q_i$
and angular momentum $J$
\begin{subequations}
\label{4d5d}
\bea
Q_i &=& p^0\,q_i + \frac12 C_{ijk}\, p^j\,p^k\label{q4d5d}\ ,\quad\\
2 J   &=& p^0 ( p^0 q_0 + p^i q_i) + \frac13 C_{ijk}\,p^i\,p^j\,p^k\ ,
\label{j4d5d}
\eea
\end{subequations}
of the 5D black hole (or more generally black ring) 
related to the 4D black hole with charges $(p^0,
p^i,q_i,q_0)$ by the 4D-5D lift~\cite{Bena:2004tk,Gaiotto:2005xt,Bena:2005ni}. 
Indeed, it is now well-known that
a four-dimensional BPS black hole with $D6$ brane charge
$p^0\neq 0$ and arbitrary $D4,D2,D0$ brane charges $p^i,q_i,q_0$ in type IIA 
string theory compactified on $X$ may be viewed at strong coupling as a 5D
black ring carrying electric M2-brane charges $Q_i$,
M5-brane dipole moments $P^i=-p^i/p^0$ and angular momentum $J_\psi=J$, 
wound around the circle of a Taub-NUT space with NUT charge $p^0$
\cite{Bena:2004tk,Gaiotto:2005xt,Bena:2005ni}. In the absence of D4-brane 
charge, the 5D configuration reduces to a single 5D black hole
placed at the tip of the Taub-NUT space, as found in \cite{Gaiotto:2005gf}.
Indeed, with this assignment of charges it may be shown that 
the Bekenstein-Hawking entropy of the 4D and 5D black holes
agree up to the orbifold factor $1/|p^0|$~\cite{Gaiotto:2005gf,Gaiotto:2005xt}.
In the context of $\cN=2$ magic supergravities, this amounts to the identity
\be
S_{4D}= \pi\sqrt{I_4(p^I,q_I)} = \frac{2\pi}{|p^0|}
\, \sqrt{N(Q_i)- J^2} = \frac{1}{|p^0|} S_{5D} 
\ee
where $I_4$ is the quartic invariant in \eqref{i4pq} \cite{Pioline:2005vi}.
It should also be noted
that the charges $Q_i,J$ defined in \eqref{4d5d} are invariant
under the ``spectral flow'' 
\bse
\be
p^0 \to p^0\ ,\qquad p^i \to p^i + p^0 \ell^i\ ,\quad
q_i \to q_i - C_{ijk}\, p^j \ell^k - \frac{p^0}{2}\, 
C_{ijk} \ell^j \ell^k\ ,\quad
\ee
\be
q_0 \to q_0 - \ell^i q_i - \frac12\, C_{ijk}\, p^i \ell^j \ell^k 
-\frac{p_0}{3}\,C_{ijk} \,\ell^i \ell^j \ell^k\ .
\ee
\ese
which corresponds to switching on a flux on the Taub-NUT 
space \cite{Gaiotto:2005gf}.

\subsection{A 5D polarization for the topological amplitude}
The form \eqref{gnp} of the holomorphic anomaly equations suggests
introducing a new polarization where the operators $\hat Q_i$ and $\hat J$
are diagonalized. For this purpose, we note that, at the classical level,
the 5D charges $(Q_i,P^i,J)$, supplemented by an extra charge 
$p_J=1/p^0$, are obtained from $(p^I;q_I)$ via a canonical
transformation generated by
\be
\label{Scan}
S( p^0, p^i ; Q_i, J) = - \frac{N(p^i)}{p^0} + Q_i \frac{p^i}{p^0} 
 - \frac{2J}{p^0}\ .
\ee
Indeed, a straightforward computation making use of the homogeneity of 
$N$ shows that 
\be
q_I = \frac{\pa S}{\pa p^I}\ ,\quad 
P^i = - \frac{\pa S}{\pa Q_i}=-\frac{p^i}{p^0}\ ,\quad
p_J = - \frac{\pa S}{\pa J} = \frac{2}{p^0}
\ee
so that 
\be
dS = q_I\,dp^I - ( P^i\,dQ_i + p_J\,dJ)\ .
\ee
This ensures that the change of variables from $(p^I;q_I)$ to 
$(Q^i,J;P_i,p_J)$ preserves the Darboux form of $\omega$,
\be
\omega = dp^I \wedge dq_I = dQ_i \wedge dP^i + dJ \wedge dp_{J}\ .
\ee
Quantum mechanically, the wave function $\Psi_{5D}(Q_i,J)$ in the 
``5D'' polarization 
where $\hat Q_i$ and $\hat J$ are diagonalized is therefore related to the
wave function $\Psi_{\IR}(p^I)$ in the ``real'' 
polarization \cite{Verlinde:2004ck}, where $\hat p^I$ acts diagonally, via
\be
\label{intert0}
\Psi_{\IR}(p^I) = \int \exp\left(-\frac{\I}{\hbar} \,S(p^I;Q_i,J) \right)\,
\Psi_{5D}(Q_i,J)\,dQ^i\,dJ\ .
\ee
Equivalently,
\be
\Psi_{\IR}(p^I)\, \exp\left( -\frac{\I}{\hbar}\, \frac{N(p^I)}{p^0} \right) = 
\int\,\exp\left( \frac{\I}{\hbar} \frac{2J}{p^0} - \frac{\I}{\hbar} 
\, \frac{p^i}{p^0} \, Q_i\right)\,
\Psi_{5D}(Q_i,J)\,dQ^i\,dJ
\label{intert}
\ee
Indeed, one may check that the operators
\be
\hat Q_i \equiv \I\hbar\,p^0\,\dwrt{p^i}+\half C_{ijk} p^j p^k \ ,\qquad
2\hat J = \I\hbar\,(p^0)^2 \dwrt{p^0} + 
\I\hbar\,p^0\,p^i\,\dwrt{p^i} + 2\,N(p^i)
\ee
acting on the l.h.s. of \eqref{intert0} lead to insertions of 
$Q_i$ and $2J$ in the integral on the r.h.s, respectively. In words,
we have found that the wave function in 
the 5D polarization is obtained by Fourier transforming the wave function
in the real polarization with respect to $1/p^0$ and $p^i/p^0$, after 
multiplication by the tree-level part $e^{-\frac{\I}{\hbar} N(p^i)/p^0}$. 

\subsection{5D polarization and 5D black hole degeneracies}

In order to interpret the result \eqref{intert}, we now recall some facts and
conjectures on the relation between the topological string amplitude and 
various invariants of the Calabi-Yau~$X$. 

First, recall that the real polarized topological wave function 
$\Psi_{\IR}(p^I)$ is related to the holomorphic topological wave function
via~\cite{Schwarz:2006br}
\be
\label{psiholR}
e^{F_{\rm hol}(t^i,\lambda)} = \left(p^0\right)^{\frac{\chi}{24}-1}\,
\Psi_{\IR}(p^I)\ ,\quad
\lambda = \frac{4\pi}{\I p^0}\ ,\quad t^i = \frac{p^i}{p^0}\ .
\ee
where $F_{\rm hol}(t^i,\lambda)$ is the holomorphic
limit $\bar t^{\bar i}\to \infty$ of the topological 
partition function $F(t^i,\bar t^{\bar i},\lambda)$.

Second, recall that the Gopakumar-Vafa conjecture~\cite{Gopakumar:1998ii,
Katz:1999xq} relates the indexed partition function of 5D spinning 
BPS black holes to the topological amplitude\footnote{Here and below,
we follow the conventions in \cite{Denef:2007vg}, up to minor changes 
of notation $g_{\rm top}\to\lambda, n\to 2J, \beta_i\to Q_i$.},
\be
\label{gvc}
e^{F_{\rm hol}(t^i,\lambda)-F_{0}(t^i,\lambda)} = 
 \sum_{Q_i,J}\, \Omega_{\rm 5D}(Q_i,J)\,
e^{-2\lambda J +2 \pi i Q_i t^i}\ .
\ee
 The conjecture also 
includes a relation to the BPS invariants $n_Q^g$ 
of the Calabi-Yau $X$~\cite{Gopakumar:1998ii},
\be
\begin{split}
\label{gvbps}
e^{F_{\rm hol}(t^i,\lambda)-F_{\rm pol}(t^i,\lambda)} 
=& [M(e^{-\lambda})]^{\chi/2}\, 
\prod_{Q_i>0,k>0} \left(1-e^{-k\lambda+2\pi\I Q_i t^i}\right)^{k n_Q^0}\\
&\times \prod_{Q_i>0,g>0}
\prod_{\ell=0}^{2g-2} 
\left(1-e^{-(g-\ell-1)\lambda+2\pi\I Q_i t^i}
\right)^{(-1)^{g+\ell} {\scriptsize \begin{pmatrix} 2g-2 \\ \ell \end{pmatrix}}
 n_Q^g}
\end{split}
\ee
Here,
\be 
F_{\rm pol}(t^i,\lambda)=-\frac{(2\pi \I)^3}{\lambda^2} N(t^i)
-\frac{2\pi\I}{24} c_{i} t^i
\ee 
is the ``polar part'' of $F_{\rm hol}(t^i,\lambda)$, and 
$M(q)=\prod(1-q^n)^{-n}$ is the Mac-Mahon function.
Unfortunately, both the BPS invariants $n_Q^g$ and 
the 5D black hole degeneracies $\Omega_{\rm 5D}(Q_i,J)$ 
so far lack a proper mathematical definition. This is in contrast to the 
now well-established relation between Gromov-Witten and 
Donaldson-Thomas invariants~\cite{Iqbal:2003ds,gw-dt}, 
\be
\label{gvc2}
e^{F_{\rm hol}(t^i,\lambda)-F_{\rm pol}(t^i,\lambda)} = 
[M(e^{-\lambda})]^{-\chi/2}\, \sum_{Q_i,J}\, (-1)^{2J}\,N_{DT}(Q_i,2J)\,
e^{-2\lambda J +2 \pi i Q_i t^i}
\ee
where $N_{DT}(Q_i,2J)$ are
the Donaldson-Thomas invariants. Physically, the latter count 
the bound states of
one D6-brane with $2J$ D0-branes and $Q_i$ D2-branes wrapped along the
$i$-th cycle in $H^{\rm even}(X,\IR)$. 

Finally, in \cite{Dijkgraaf:2006um}, the 4D-5D lift was used  to argue that
$N_{DT}(Q_i,2J)\sim\Omega_{\rm 5D}(Q_i,J)$, thereby giving a 
heuristic derivation of the Gopakumar-Vafa conjecture \eqref{gvc}.
However, this argument does not account for the powers of the 
Mac-Mahon function in \eqref{gvc2} relative to \eqref{gvc}, nor
for the sign $(-1)^{2J}$. There is also a discrepancy 
(most likely due to a difference in the treatment of
the center of motion degrees of freedom) between 
the prediction of the infinite product representation \eqref{gvbps},
$N_{DT}(Q,2J)=\sum_{g} \scriptsize \begin{pmatrix} 2g-2 \\ 2J+g-1\end{pmatrix}
n_Q^g$, and the considerations in \cite{Katz:1999xq,Huang:2007sb}, which 
lead to $\Omega_{5D}(Q,J)=\sum_{g} \scriptsize
\begin{pmatrix} 2g+2 \\ 2J+g+1\end{pmatrix}
n_Q^g$. Without attempting to resolve these issues, we 
shall regard \eqref{gvc} as a definition of the 5D black hole degeneracies 
$\Omega_{\rm 5D}(Q_i,J)$, and later assume that $\log\Omega_{\rm 5D}(Q_i,J)$
is given by the Bekenstein-Hawking-Wald formula, barring any "miraculous" 
cancellations.

Substituting \eqref{psiholR} into \eqref{gvc} and setting $c_i=0$
for simplicity, we obtain
\be
\label{intert2}
\left(p^0\right)^{\frac{\chi}{24}-1}\,
\Psi_{\IR}(p^I)\, \exp\left( \frac{\I\pi}{2}\, \frac{N(p^I)}{p^0}\right) = 
\sum_{Q_i,J}\,\exp\left( 8\pi\I \frac{J}{p^0} 
+ 2\pi\I Q_i \,\frac{p^i}{p^0} \right)\,\Omega_{5D}(Q_i,J)\ .
\ee
Barring the power of $p^0$, allowing for rescalings of $Q_i$ and $J$
and setting the Planck constant to $\hbar=-2/\pi$, we see that 
\eqref{intert} and \eqref{intert2} are 
consistent provided the topological wave function in the 5D polarization 
has delta function support on integer charges $Q_i,J$, with weights equal
to the 5D black hole degeneracies,\footnote{The factors of $1/4$ and $-1/8$ 
in this equation are convention-dependent, and so is the value of $\hbar$.}
\be
\label{psiom}
\Psi_{5D}(Q_i,J) \sim \sum_{Q_i',J'} \Omega_{5D}(Q_i',J')\,
\delta(Q_i'-\frac14 Q_i,J'+\frac18 J)\ .
\ee
The power of $p^0$ in \eqref{intert2} may be attributed to a quantum
ordering ambiguity invisible in the semi-classical discussion in 
the previous Subsection, or may be absorbed in a redefinition of 
$\Omega_{5D}(Q_i,J)$. Note also that \eqref{intert} was motivated in
the context of magic supergravities, but that \eqref{intert2} holds
(to the same extent as \eqref{gvc}) in arbitrary $\cN=2$ string
compactifications. 

Thus, we conclude that the 5D black hole degeneracies $\Omega_{5D}(Q_i,J)$ 
can be viewed as a wave function in a particular ``5D'' polarization, 
related to the standard real polarization by the intertwining 
operator \eqref{intert2}. 
The fact that
the degeneracies $\Omega_{5D}(Q_i,J)$ can be interpreted as 
components of a wave 
function in a representation space of the group $\tilde G$ gives
some support to the general expectation (voiced 
e.g. in \cite{Pioline:2005vi,Gunaydin:2005mx})
that they should arise as Fourier coefficients
of a certain automorphic form of $\tilde G$.

\subsection{Black hole entropy and asymptotics of the topological
amplitude}

Assuming the validity of the Gopakumar-Vafa conjecture \eqref{gvc} 
(and regardless of the correctness
of the identification \eqref{psiom}), we can
use our knowledge of the entropy of 5D black 
holes to constrain the asymptotic 
behavior of the topological string amplitude. Recall that 
the Bekenstein-Hawking entropy
of 5D BPS spinning black holes is given at tree level 
by \cite{Breckenridge:1996is}
\be
\label{S5D}
S_{5D} = 2\pi\,\sqrt{Q^3 - J^2} 
\ee
where $Q$ is to be expressed in terms of the electric charges via
\be
Q^{3/2} = \frac16 C_{ijk} Q^i Q^j Q^k\ ,\quad Q_i = \frac12 C_{ijk} Q^j Q^k
\ee
Equation \eqref{S5D} is valid in the limit where $Q_i$ and $J$ are scaled to 
infinity, keeping the ratio $J^2/Q^3$ fixed and less than unity.
Taking into account higher-derivative corrections of the 
form $\int c_i \, A^i
\wedge R \wedge R$ together with their supersymmetric partners, 
the Bekenstein-Hawking-Wald entropy becomes  \cite{Castro:2007ci}
\be
\label{S5Dcor}
S_{5D} = 2\pi\,\sqrt{Q^3 - J^2} 
\, \left( 1 + \frac{c_i \,Q^i}{16} \,\frac{Q^{3/2}}{Q^3 - J^2}\,
+ \mathcal{O}(c^2)  \right)
\ee
which is valid in the same regime.
The free energy of rotating BPS black holes in 5 dimensions 
in a thermodynamical ensemble with electric potentials $\phi^i$ and 
angular velocity~$\omega$,
\be
\label{f5leg}
F_{5D}(\phi^i,\omega) \equiv {\rm Extr}_{Q_i,J}\,\left[
S_{5D} - \omega J - Q_i \phi^i \right] 
\ee
is easily computed to first order in $c_i$, 
(see Appendix B for details)
\be
F_{5D}(\phi^i,\omega) = 
-\frac{1}{\pi^2}\,\frac{N(\phi^i)}{1+(\omega/2\pi)^2} - 
\frac18 c_i \phi^i + \mathcal{O}(c^2)\ .
\label{f5exp}
\ee
This results holds for arbitrary, non-magic supergravities in 5D,
and is considerably more elegant than its Legendre 
dual \eqref{S5Dcor}\footnote{The
apparent discrepancy at order $c_i$ with the free energy given in Eq. (3.23) of
\cite{Castro:2007ci} is due to the fact that the chemical potentials
$e_I$ are conjugate to the 4D charges rather than the ones measured
at infinity in 5D, as recognized in \cite{Castro:2007ci}. At zeroth
order in $c_i$, the result \eqref{f5exp} was known to the second author,
R.~Dijkgraaf and E.~Verlinde in 2005.}.

The free energy \eqref{f5exp} provides the classical (saddle point) 
approximation to 
the integral in \eqref{intert}. The fluctuation
determinant around the saddle point may be computed in the magic
cases using the results in \cite{Pioline:2005vi}. Setting
\be
\phi^i = -2\pi \I\, t^i\ ,\quad \omega = 2 \lambda\ ,
\ee
we find
\be
\label{lim5}
\Psi_{\IR}(X^I) \sim \lambda^{\frac{\chi}{24}-1}\,
\left( \frac{N(t_i)}{\lambda^2+\pi^2} \right)^{\frac{n_v+3}{6}}\,
\exp\left[ -(2\pi i)^3 \left( \frac{N(t^i)}{\lambda^2} 
-  \frac{N(t^i)}{\pi^2+\lambda^2} 
\right)+\frac{2\pi i}{8} c_i t^i + \dots  \right]  
\ee
where the prefactor can be trusted in magic cases only.
In the scaling limit where the Bekenstein-Hawking formula can
be trusted, the topological coupling $\lambda$ at the saddle point is fixed 
while $t^i$ are 
scaled to infinity, so that the terms displayed in \eqref{f5exp} are
the first two in a systematic expansion at large $t^i$, for fixed
$\lambda$. It is noteworthy that the terms proportional to $1/\lambda^2$ and 
$1/(\pi^2+\lambda^2)$ in the exponent cancel in the limit of large $\lambda$, 
leaving a term of order $1/\lambda^4$ only. Incidentally, we note that the 
linear term in $t^i$ in the exponent induces a correction $Q_i\to
Q_i-\frac18 c_i$ to the 4D-5D lift  formulae \eqref{4d5d}, consistent
with \cite{Castro:2007hc,Castro:2007ci} in the absence of 
angular momentum, but giving a different correction than the one
found in \cite{Castro:2007ci} when $J\neq 0$.

On the other hand, at small topological coupling and finite values 
of $t^i$, \eqref{psiholR} yields
\be
\label{lim4}
\Psi_{\IR}(p^I) \sim (p^0)^{1-\frac{\chi}{24}}
\exp\left[-\frac{\I\pi}{2} F_0(p^I)\right] \sim 
\lambda^{\frac{\chi}{24}-1}\,
\exp\left[ -(2\pi\I)^3\frac{N(t^i)}{\lambda^2} 
-\frac{2\pi\I}{24} c_i t^i +\dots\right]
\ee 
The semi-classical limit $\lambda\to 0$ at fixed $t^i$ is consistent 
with the entropy of 4D BPS black holes, a fact which lies at the basis of the 
OSV conjecture \cite{Ooguri:2004zv}.
For completeness, we show in Appendix B how \eqref{lim4} is consistent with the
usual form of the BCOV topological amplitude,
\be
\label{bcovexp}
\Psi_{\rm BCOV} (t_i,\bar t_i,x^i,\lambda) \sim \lambda^{\frac{\chi}{24}-1}
\,  \exp\left( -(2\pi\I)^3 \frac{C_{ijk} x^i x^j x^k}{\lambda^2} +
\mathcal{O}(\lambda^0) \right)
\ee
after performing the sequence of transformations given 
in \cite{Gunaydin:2006bz}.

The regimes of validity of  \eqref{lim5} and \eqref{lim4}
in principle overlap when $\lambda$ goes to zero and $t^i$ to infinity. 
While the two results agree in the strict classical limit,
the prefactors do not. Moreover matching the terms in the
exponent would require $F_1(t^i)\sim \frac{2\pi\I}{8}
c_i t^i + \frac{(2\pi\I)^3}{\pi^2}N(t^i)$, which 
violates the assumption that $F_1$ is grows linearly 
at large $t^i$. This discrepancy suggests that the two limits
$t^i\to \infty$ and $\lambda$ do not commute. It would be interesting
to understand the physical origin of this phenomenon.

\section{The generalized topological amplitude in the 5D polarization}

In the previous section, we have shown that the 4D-5D lift formula
and Gopakumar-Vafa conjectures had a simple interpretation as a change
of polarization in the Schr\"odinger-Weil representation of 
the Fourier-Jacobi group $\tilde G = G \ltimes H$, where $G$ is 
the four-dimensional U-duality group and $H$ is the Heisenberg algebra 
of electric, magnetic and NUT charges. 

\subsection{Dimensional reduction and extended topological amplitude}

Physically, the Fourier-Jacobi group $\tilde G$ naturally arises
as a subgroup of a larger group $G'$, the duality group after
reducing the 4D supergravity  (or compactifying) down to 3 dimensions.
After dualizing the one-forms into scalars, the theory in 3 dimensions
can be expressed as a non-linear sigma model on a quaternionic-K\"ahler
space $\mathcal{M}_3 = G'/SU(2)\times G_c$, where $G_c$ is a compact
form of the duality group $G$ in 4 dimensions~\cite{Breitenlohner:1987dg}. 
The reduction from $G'$ 
to its subgroup $\tilde G$ corresponds to decoupling gravity in 3
dimensions. In the language of Jordan algebras, $G'={\rm QConf}(J)$
is the ``quasi-conformal group'' associated to the Jordan 
algebra $J$ \cite{Gunaydin:2000xr,Gunaydin:2007qq,Pioline:2006ni}.
Its Lie algebra can be obtained by supplementing the solvable 
group $\tilde G$ with the negative roots $\hp^{I'}, \hq_I, Z'$,
obeying the ``dual Heisenberg algebra'' 
$[\hp^{I'}, \hq_J^{'}]= Z'\, \delta^I_J$,
and introducing a new Cartan generator $\Delta \equiv [Z,Z']$,
such that $Z,Z',\Delta$ forms a $Sl(2,\IR)$ subalgebra commuting with $G$
(see Figure \ref{qconfr}). 

\EPSFIGURE{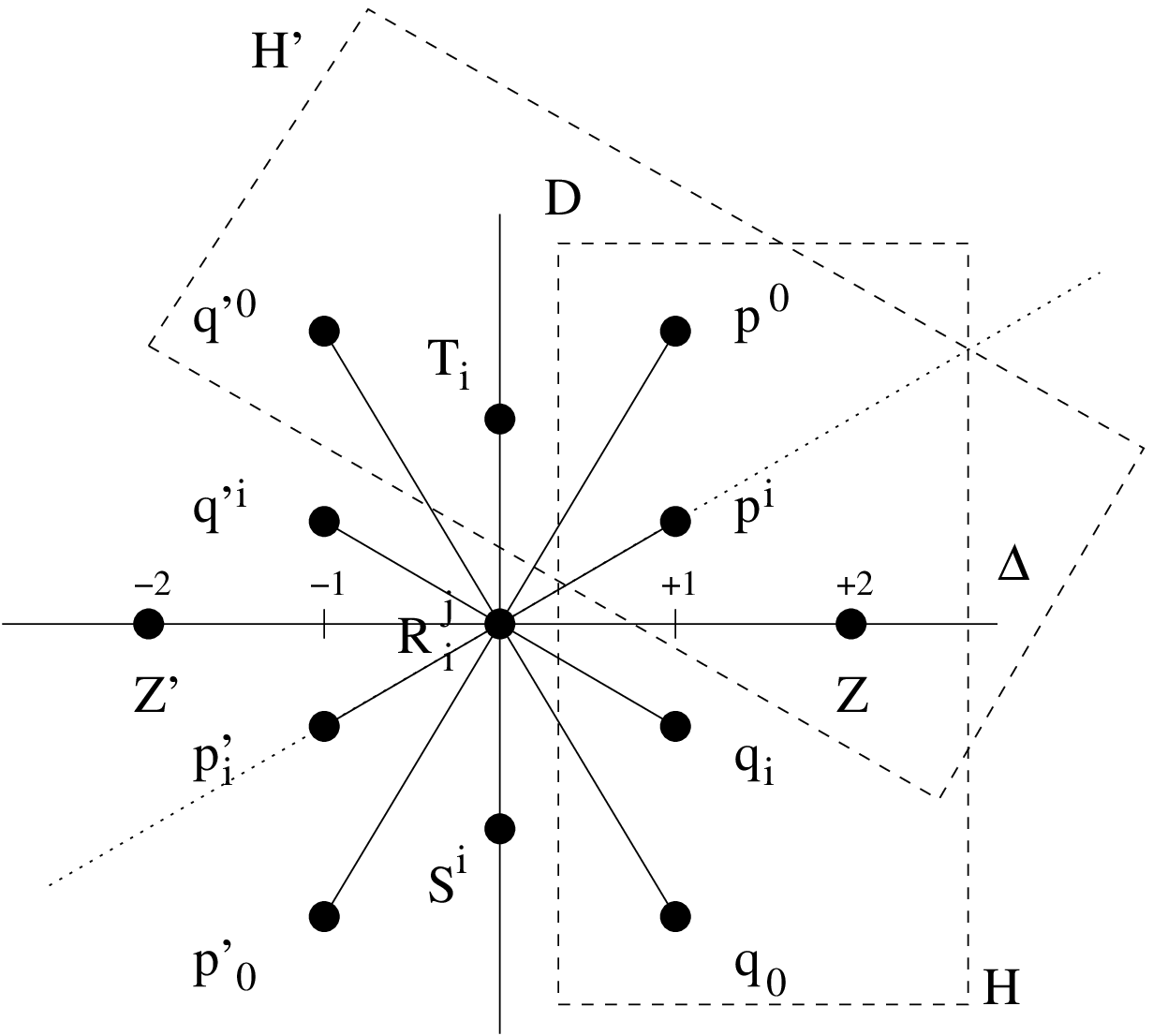, height=10cm}{Two-dimensional projection of the
root diagram of $G'$, with respect to the split torus $(\Delta,D)$.
The subgroup $G$ consists of the roots along the vertical axis. 
The Fourier-Jacobi subgroup $\tilde G=G\ltimes H$ consists 
of the roots on and to right of the vertical axis, together with 
the Cartan generator $\Delta$. The Heisenberg 
algebras $H$ and $H'$ are exchanged by a Weyl reflection $W$ with respect to 
the dotted axis.
\label{qconfr}}

Moreover, the group $G'$ admits a distinguished unitary representation
known as the ``minimal'' representation, whose functional dimension $n_v+2$ 
is the smallest among the unitary irreducible representations of $G'$. 
The minimal representation of $G'$ extends the Schr\"odinger-Weil 
representation of $\tilde G$ in the following way: classically,
the Freudenthal triple \eqref{RJJR} is extended into
\be
\label{ERJJR}
V'= \IR_{y} \oplus \IR_{p^0} \oplus J_{p^i} \oplus 
J_{q_i} \oplus \IR_{q_0} \oplus \IR_{p_y}
\ee
equipped with the symplectic form
\be
\omega' = 
dy \wedge dp_y + dp^0 \wedge dq_0 + dp^i \wedge dq_i \ .
\ee
The linear space $V'$ turns out to be symplectically isomorphic to the minimal
co-adjoint orbit of the complexification of $G'$ (itself isomorphic
to the hyperk\"ahler cone over the quaternionic-K\"ahler space 
$G'/(SU(2)\times G_c)$), and therefore 
admits a  holomorphic symplectic action of $G'$ on 
\eqref{ERJJR}. The minimal representation is obtained by quantizing this 
action (see 
e.g. \cite{Gunaydin:2007qq} for details on this procedure).
Quantum mechanically, the minimal representation of $G'$ may be obtained 
from the Schr\"odinger-Weil representation \eqref{swrep} by allowing 
the center $Z=\I\hbar$ to become dynamical, i.e. supplement the 
Hilbert space $\mathcal{H}$ of $L^2$ functions of $n_v+1$ variables $p^I$ 
with an extra variable $y$, and set
\footnote{With this 
notation the scalar $p^I$ differs from the 
eigenvalue of $\hp^I$ by a power of $y$. We hope that this will not cause
any confusion.},
\bse
\begin{gather}
Z \mapsto \I y^2, \label{minrep-1} \\
\I\hq_0 \mapsto y \dwrt{p^0}, \quad 
\I\hq_i \mapsto y \dwrt{p^i}, \quad \I\hp^i \mapsto \I y p^i, \quad \I
\hp^0 \mapsto \I y p^0,   \label{minrep-2} 
\end{gather}
while keeping the same formulae for the action of $G$ 
as in \eqref{swrep-5}. The rest of the generators of $G'$ are obtained by
commuting the generators above with 
\be
Z' \mapsto \half \frac{\pa^2}{\pa y^2} - \frac{1}{4 y^6} \left( 
I_4(\hp^I, \hq_I) + \kappa \right)\ ,\quad
\Delta \mapsto y \pa_y + \frac12 \label{minrep-3} 
\ee
where the constant $\kappa$ depends on the ordering chosen 
in $I_4(\hat p^i,\hat q_i)$. In particular,
\bea
\I{\hq}_I'
&\equiv& [\hq_I,Z'] \mapsto \I \dwrt{p^I} \pa_y + \frac{1}{y^4}
\frac{\pa I_4(\hat p^I,\hat q_I)}{\pa \hat p^I}\ ,\label{minrep-4} \\
\I{\hp}'^{I}
&\equiv& [\hp^I,Z'] 
\mapsto \I p^I \pa_y - \frac{1}{y^4} 
\frac{\pa I_4(\hat p^I,\hat q_I)}{\pa \hat q_I} \label{minrep-5} 
\ .
\eea
\ese
These formulae define the minimal representation 
in the real polarization, where the operators 
$\hp^I$ and $Z$ are diagonalized. At fixed value of $y$, 
the representation of the subgroup $\tilde G\subset G'$ 
reduces to the Schr\"odinger-Weil 
representation studied in the previous section,
after appropriate $y$-rescalings.

As argued in \cite{Gunaydin:2006bz}, the relation 
between $\tilde G$ and $G'$ on the one hand,
and between the Schr\"odinger-Weil 
representation of $\tilde G$ and the  minimal representation
of $G'$ on the other hand, is closely analogous to the relation of 
the Fourier-Jacobi group $Sl(2,\IR)\ltimes H_3$ and
Siegel's genus 2 modular group $Sp(4,\IR)$, familiar from the 
mathematical theory of Jacobi and Siegel modular 
forms~\cite{eichlerzagier} (Here $H_3$ is the three-dimensional
Heisenberg algebra $[p,q]=Z$, where $(p,q)$ transform as a doublet
of $Sl(2,\IR)$). In that case, the
Schr\"odinger-Weil representation of $Sl(2)\ltimes H_3$ on $L^2$-functions
of one variable is then given by the restriction of the metaplectic 
representation of  $Sp(4,\IR)$ on $L^2$-functions of two variables, 
at a fixed value of the center $Z$. At the automorphic level, the $m$-th
Fourier coefficient
of a Siegel modular form with respect to the action of the center $Z$ yields 
a Jacobi form of $Sl(2,\IZ)\times H_3$ of index $m$~\cite{eichlerzagier}.
Based on this analogy, it was suggested in \cite{Gunaydin:2006bz} that,
in cases where the vector multiplet moduli space is symmetric, 
the standard BCOV topological 
amplitude should arise as a Fourier coefficient
at $Z=-\I$ of an automorphic form
under the larger group $G'={\rm QConf}(J)$, referred
to as the ``extended topological amplitude''.
It was further speculated in \cite{Gunaydin:2006bz} that the Fourier 
coefficients at other values
of $Z$ yield non-Abelian generalizations of the Donaldson-Thomas
invariants.

At this point, we note that the dimensional reduction to 3 dimensions,
which has been of great utility in describing four-dimensional 
stationary black holes \cite{Breitenlohner:1987dg,Gunaydin:2005mx,
Gunaydin:2007bg}, is also very useful in order to describe five-dimensional 
black holes with a $U(1)$ isometry~\cite{Maison:1979kx,Giusto:2007fx,
Bouchareb:2007ax,Berkooz:2008rj}. The two reductions differ, however, 
since 5D black holes are best described by reducing the 5D Lagrangian
along the time-like direction $t$ first, and then along a space-like direction
$\psi$, while 4D black holes are more conveniently described by first reducing
from 5D to 4D along the space-like direction $\psi$, and then 
from 4D to 3D along the time-like direction $t$. The two procedures are 
related by a Weyl reflection $W$ inside the diffeomorphism group 
of the $(t,\psi)$ torus, which happens to be the $Sl(2)$ subgroup
of $G'$ generated by $\hq_0, \hq_0^{'}$ and their 
commutator \cite{Berkooz:2008rj}. 
The Weyl reflection $W$ maps the Heisenberg algebra $\{p^I,q_I,Z\}$
(enclosed in the vertical box of Figure \ref{qconfr}) to the Heisenberg algebra
$H'=\{ \hq_0', T_i, \wh{p}^i, Z, \wh{p}^0 \}$ (enclosed by the tilted
box). In particular, the D2 and D0 brane charges 
$\hq_i$ and $\hq_0$ are mapped to $T_i$ and $\hq_0^{'}$. According 
to \eqref{gnp} and \eqref{minrep-4} above, the corresponding generators 
in the minimal representation are given by 
\bea
\label{tqp0min}
\I\,T_i = \frac{1}{y^2} \left( 
\hp^0 \hq_i + \frac12 C_{ijk} \hp^j \hp^k\right)\ ,\quad
\hq_0' = \frac{1}{y^4} \left[ 
\hp^0 (\hp^0 \hq_0 + \hp^i \hq_i) + 2 N(\hp) \right]
+ \frac{1}{2y} \hp^0 \hp_y
\eea
where $\hp_y=\I \pa_y$.
This are indeed  the 5D
electric charges $Q_i$ and angular momentum $J$ in \eqref{4d5d}, 
up to a normalization factor and
and additive term in $\hq_0^{'}$\footnote{Despite the fact that 
equations \eqref{tqp0min} 
hold only in the minimal representation, whose semi-classical
limit pertains to special solutions whose Noether charge is nilpotent of 
degree 2, the equality of the conserved charges $(T_i,\hq_0')$ with 
the electric charge and angular momentum holds in 
general \cite{Berkooz:2008rj}
.}. Moreover, the unit D6-brane charge requirement $p^0=1$, appropriate
for lifting a 4D black hole to a smooth 5D black hole, 
is mapped to $Z=-\I$, which is the necessary requirement for the $\psi$
circle bundle over $S^2$ to be topologically $S^3$  \cite{Berkooz:2008rj}. 
Conversely, the vanishing of the time-like NUT charge $Z=0$ 
for 4D black holes is mapped to the absence
of $p^0$ charge for 5D black holes~ \cite{Berkooz:2008rj}. 

\subsection{A 5D polarization for the minimal representation}

We now construct the analogue of the 5D polarization in
this generalized setting. For this purpose, we need to
supplement the 5D charges $(Q_i,J)$ and their canonical conjugate
$(P^i,p_J)$ with an extra canonical pair $(L,p_L)$, preserving
the fact that $(Q_i,J)$ are related to $(p^I,y,q_I,p_y)$ by 
\eqref{tqp0min}. The canonical transformation generated by 
\be
\label{genfg}
S'(p^0,p^i,p_y;Q_i,J,L) = \frac{N(p^i)}{p^0} - \frac{p^i}{p^0}\,Q_i  +  
\frac{2J L}{(p^0)^2} + L \frac{p_y}{p^0} 
\ee
satisfies these conditions (compare to \eqref{Scan}). Indeed,
after some algebra, one finds that the 5D phase space variables
$(Q_i,J,L;P^i,p_J,p_L)$ are related to the 4D phase space variables
$(p^I,p_y;q_I,y)$ via
\bse
\label{q4d5dg}
\bea
Q_i &=& p^0\,q_i + \frac12 C_{ijk}\, p^j\,p^k\ ,\\
2J   &=& \frac{1}{y} \left[ 
p^0 ( p^0 q_0 + p^i q_i) + \frac13 C_{ijk}\,p^i\,p^j\,p^k \right]
+ \frac12 p^0 p_y\\
L   &=& p^0 y \ ,\quad
P^i = \frac{p^i}{p^0}\ ,\quad p_J= \frac{y}{p^0}\ ,\quad \label{q4d5dg3}\\
p_L &=& \frac{1}{y\, (p^0)^2} \left[ 
p^0 ( p^0 q_0 + p^i q_i) + \frac13 C_{ijk}\,p^i\,p^j\,p^k \right]
- \frac{p_y}{2 p^0}
\eea
\ese
The generating function of the canonical transformation from $(p^0,p^i,y)$ 
to $(Q_i,J,L)$ is obtained by Legendre transforming \eqref{genfg} with 
respect to $p_y$, which removes the last term in \eqref{genfg} 
and sets $y=L/p^0$ consistently with \eqref{q4d5dg3} above.
Quantum mechanically, the wave function in the generalized 5D 
polarization $\Psi_{5D}(Q^i,J,L)$ is therefore related to the 
generalized wave function in the real polarization $\Psi_{\rm gen}(p^0,p^i,y)$ 
via
\be
\Psi_{\rm gen}(p^I,y)\, e^{-\I\, N(p^i)/p^0} = 
\int\,\exp\left( 2\I \frac{y\,J}{p^0} - \I \, \frac{p^i}{p^0} \, Q_i\right)\,
\Psi_{5D}(Q_i,J,L)\,\delta(L-p^0 y)\,dQ^i\,dJ\,dL
\label{intertg}\ .
\ee
In this new polarization, the ``tilted'' Heisenberg algebra $H'$, with 
center $\wh{p}^0$ is now canonically represented,
\be
\begin{matrix}  \hq_0' = 2\I J\ &,& \quad T_i = \I\,Q_i\ ,\\
Z' = \frac12 L \pa_J\ &,& \quad \wh{p}^i = L \pa_{Q_i}\ ,
\end{matrix}
\quad  \wh{p}^0 = \I L\ .
\ee
In fact, the intertwiner \eqref{intertg} represents the 
action of the Weyl reflection $W$, which takes $H$ into $H'$. Thus, 
all generators in the 5D polarization 
can be obtained from those in the 4D polarization by reflecting
the root diagram in Figure 1 along the dotted axis and changing variables
\be
2J \to p^0, \quad Q_i \to y p^i\ ,\quad L \to y^2 \ .
\ee
For example,
\be
\label{ZDEZ}
Z= \frac12 L \pa_J\ ,\quad \Delta = L \pa_L - J \pa_J\ ,\quad 
Z'= 2J \pa_L + \frac{2J}{L} - \frac{N(Q_i)}{L^2}\ .
\ee
These results agree and generalize the ones obtained for $G'=G_{2(2)}$
in Section 3.7.2 of \cite{Gunaydin:2007qq}, after performing an overall
Fourier transform over all $p_I$. Note that \eqref{ZDEZ}
implies that $(2J,L)$ transform linearly as a doublet under the $Sl(2,\IR)$ 
symmetry generated by $Z,Z',\Delta$, which is Ehlers' symmetry 
in four dimensions. Thus, the 5D polarization
constructed here would be the most convenient starting point 
to implement Ehlers' symmetry on the generalized topological string 
amplitude.

\section{Discussion}
In this note, motivated by the formal analogy between the holomorphic
anomaly equations and the  4D-5D lift formulae for ``magic'' supergravities,
we gave a quantum mechanical interpretation of the Gopakumar-Vafa relation
as a Bogoliubov transformation from the real polarization, where the
4D magnetic charge operators $\hp^I$ operators act diagonally, to the
the ``5D'' polarization, appropriate to the operators $\hat Q_i$ and 
$\hat J$. Moreover, we used to the known Bekenstein-Hawking-Wald
entropy of 5D BPS black holes to constrain the asymptotic behavior of the 
topological wave function in the real polarization, at finite topological
coupling but large K\"ahler (or, in the B-model, complex structure) 
moduli $t^i$. 

In the process we found two relatively minor discrepancies: (i) 
a yet unexplained shift of genus $g\to g-2$ in the relation between 
$N_{DT}(Q_i,2J)$ and $\Omega_{\rm 5D}(Q_i,J)$, 
and (ii) a disagreement
at subleading order in the expected overlapping regime of validity of the
asymptotic expansions afforded by the 4D and 5D black hole entropy.
The former may probably be solved by a proper accounting of the 
zero-modes of a 5D black hole at the tip of Taub-NUT space, while the
latter suggests a non-commutativity of the limits $\lambda\to 0$
and $t^i\to\infty$.
It would certainly be useful to resolve these puzzles, and improve
our understanding of 5D black hole micro-states. 

In the last section of this paper, we extended the construction of
the 5D polarization to the case where gravity is no longer decoupled,
and the duality group is enlarged from $G\ltimes H$ to a semi-simple 
Lie group $G'$. In particular, we found that the intertwiner from the
real to the 5D polarization represents a particular Weyl reflection in
the 3D duality group $G'$, which exchanges the two directions in the
internal 2-torus. Assuming that a ``generalized topological amplitude''
living in the minimal representation of $G'$ can really be defined, 
it is interesting to ask what information it may capture. In 
\cite{Gunaydin:2006bz}, it was suggested that $\Psi_{\rm gen}
(p^I,y)$ would
give access to non-Abelian Donaldson-Thomas invariants of rank $y^2$. 
The 5D polarized
wave function $\Psi_{5D}(Q_i,J,L)$
constructed herein naturally suggests an interpretation in terms of 
counting 5D black hole micro-states of charge $Q_i$, angular momentum $J$
and dipole charge $p^0 \propto L$. It would be interesting to see if 
this conjecture can be borne out.

\acknowledgments

We are grateful to I.~Bena, M.~Berkooz, A.~Dabholkar, 
R.~Dijkgraaf,
P.~Kraus, K.~Hori, D.~Jafferis, S.~Murthy,  
A.~Neitzke and T.~Pantev for useful discussions. 
P.G. is supported in part by NSERC.
B.P. is supported in part by ANR (CNRS--USAR) contract No 05--BLAN--0079--01.

\appendix

\section{Free energy of 5D spinning black holes}
In this appendix, we derive Eq.~\eqref{f5exp} for the free energy
of 5D spinning black holes. We start by computing the Legendre
transform of the tree-level entropy \eqref{S5D}, and incorporate
the higher-derivative corrections at the end.

Extremizing \eqref{f5leg} over $J$, we find that the extremum
is reached at 
\be
J=-\frac{\omega}{\sqrt{\omega^2+4\pi^2}}\,Q^{3/2}
\ee
leaving 
\be
F_{5D}(\phi^i,\omega) = 
\langle \sqrt{\omega^2+4\pi^2}\, 
Q^{3/2}  - Q_i \phi^i \, \rangle_{Q_i}\ .
\ee
The extremum over $Q_i$ is therefore reached at
\be
Q^i = \frac{1}{\pi} \frac{\phi^i}{\sqrt{1+(\omega/2\pi)^2}}\ ,
\ee
at which point
\be
F_{5D}(\phi^i,\omega) = -\frac{1}{\pi^2} \frac{N(\phi^i)}{1+(\omega/2\pi)^2}
\ee
To incorporate the effect of the higher-derivative correction
in \eqref{S5Dcor}, we note that the variation of the tree-level
entropy \eqref{S5D} with respect to $q_i$ is given, to leading order,
by 
\be
\delta S_{5D} = \frac{\pi Q^{3/2}}{\sqrt{Q^3-J^2}} Q^i \delta Q_i 
\ee
where we used the fact that $\delta Q^{3/2} = \frac12 Q^i \delta Q_i$.
Thus, the subleading term in \eqref{S5Dcor} is reproduced by setting 
$\delta Q_i = \frac18 c_i$. After Legendre transform, the
corrected free energy is therefore
\be
F_{5D}(\phi^i,\omega) = -\frac{1}{\pi^2} \frac{N(\phi^i)}{1+(\omega/2\pi)^2}
 - \frac18 \phi^i c_i + \dots
\ee
Upon scaling $Q_i$ and $J$ to infinity keeping $J/Q^{3/2}$ fixed and less than
one, it is easy to see that $\omega$ is fixed while $\phi^i$ go to infinity.
The limit $\omega\to\infty$ (corresponding to strong topological coupling)
corresponds to black holes near the Kerr bound $J=Q^{3/2}$.

\section{From BCOV to real polarization}
In this appendix, we provide a check on \eqref{psiholR} in the case
of ``magic'' $\cN=2$ supergravities, which illuminates the
relation between the constructions in \cite{Gunaydin:2006bz} and
\cite{Schwarz:2006br}. For this purpose, we postulate the form
\be
\label{eguess}
\Psi_{\IR}(p^I) \sim 
(p^0)^{\alpha}\,\exp\left[ -\frac{\I \pi}{2}\,\frac{N(p^i)}{p^0} \right]\ ,
\ee
consistent with \eqref{psiholR} for $\alpha = 1 - \frac{\chi}{24}$,
and show that it leads to a BCOV topological partition function 
of the expected form \eqref{bcovexp} after applying the chain of
transformations in \cite{Gunaydin:2006bz}. 
The first step is to obtain the holomorphic wave function 
via~\cite{Gunaydin:2006bz}
\be
\label{interhol}
\Psi_{\rm hol}(t^i;w,y_i) = \int dp^I\,
\exp\left( \frac{\I\pi}{4} p^I \tau_{IJ}(X) p^J + \frac{\pi}{2} 
p^I y_I \right)
\, \psi_{\IR}(p^I)\ .
\ee
where $t^i=X^i/X^0$ and $w=y_0+t^i y_i$.
To evaluate this integral in the saddle point approximation, 
define $\tilde p^i = p^i - p^0 X^i/X^0, \tilde p^0=p^0$. Taylor expanding
the r.h.s. at small $p^0$, it is easy to check that 
\be
 -\frac{1}{2} p^I \tau_{IJ}(X) p^J 
=\frac{N(\tilde p^i)}{\tilde p^0} - \frac{N(p^i)}{p^0}) \ .
\ee
Moreover, defining $\tilde y_0=y_0+ y_i X^i/X^0, \tilde y_i=y_i$, we have
\be
p^I y_I = \tilde p^0 \tilde y_0 + \tilde p^i y_i  \ .
\ee
Inserting \eqref{eguess} into \eqref{interhol} and changing variables 
from $p^I$ to $\tilde p^I$ leads then to 
\be
\Psi_{\rm hol}(X^I,y_I) = \int d\tilde p^I\,
(\tilde p^0)^{\alpha}\,
\exp\left[ -\frac{\I\pi}{2} \frac{N(\tilde p^i)}{\tilde p^0} +  \frac{\pi}{2} 
\tilde p^I \tilde y_I \right]\ .
\ee
In the saddle point approximation, using the results in \cite{Pioline:2005vi},
we conclude that
\be
\label{psiholap}
\Psi_{\rm hol}(X^I,y_I) \sim 
(\tilde y_0)^{\alpha'}\,[N(\tilde y_i)]^{\beta'}\,
\exp\left[\frac{\I\pi}{2} \frac{N(\tilde y_i)}{\tilde y_0} \right] \ ,
\ee
where (except in the $D_n$ case)
\be
\alpha' = -2\alpha-\frac12(n_v+3)\ ,\quad
\beta' = \alpha+\frac16(n_v+3)\ .
\ee
Next, we take the complex conjugate $\Psi_{\rm ahol}(\bar X^I, \bar y_I)$ of
\eqref{psiholap} and change variable from $\bar y_I$ to $x^i,\lambda$ using
\be
\label{xtoy}
x^I \equiv [\Im\tau]^{IJ} \bar y_J 
= 2\,e^{-\frac14 \pi \I}\, \lambda^{-1}\ (X^I + \, x^i \, D_i X^I) \ .
\ee
The  BCOV topological partition function is finally obtained as
\be
\label{holtobcov}
\Psi_{\rm BCOV}(t^i,\bar t^i,x^i,\lambda) = e^{-f_1(t)}\,
\sqrt{\det[\Im\tau]}\,\exp\left(-\pi\, x^I [\Im\tau]_{IJ} x^J\right)
\,  \Psi_{\rm ahol}(\bar X^I, \bar y_I)
\ee 
For magic supergravities, and in the gauge $X^0=\bar X^0=1$, 
equations \eqref{xtoy} are solved by
\be
\bar{\tilde y}_0 = \I\,\lambda^{-1}\,  e^{-K}\ ,\quad
\bar y_{\bar i} = -\I\, \lambda^{-1} e^{-K} g_{\bar i j} 
\left( x^j - (t^j - {\bar t}^j)\right)\ ,
\ee
Moreover,
\be
\det[\Im\tau] = e^{-\frac{n_v+3}{3}K}\ ,\quad 
N(\bar y_{\bar i}) = \I \lambda^{-3}\,e^{-K}\,N\left(x^j - (t^j - {\bar t}^j)\right)\ .
\ee
Altogether, \eqref{holtobcov} evaluates to
\be
\Psi_{\rm BCOV} (t_i,\bar t_i,x^i,\lambda) \sim 
\lambda^{-\alpha}\,\left( \frac{N(t^i-\bar t^i)}
{N\left(x^i - (t^i - {\bar t}^i)\right)}\right)^{\beta'}\,
\exp\left[ \frac{i\pi}{2} \frac{N\left(x^i - (t^i - {\bar t}^i)\right)}
{\lambda^2}-\pi x^I [\Im\tau]_{IJ} x^J\right]
\ee
The quadratic correction in the exponent cancels the terms of order 0,1,2
in $x^i$, leaving only the cubic term in the exponent,
\be
\label{finan}
\Psi_{\rm BCOV} (t_i,\bar t_i,x^i,\lambda) \sim 
\lambda^{-\alpha}\,\left( \frac{N(t^i-\bar t^i)}
{N\left(x^i - (t^i - {\bar t}^i)\right)}\right)^{\beta'}\,
\exp\left( \frac{\I \pi}{2} \frac{N(x^i)}{\lambda^2}\right)
\ee
Thus, we find agreement with the expected form
\eqref{bcovexp} provided we set 
\be
f_1(t)=0\ ,\quad \alpha = 1 - \frac{\chi}{24}
\ee
This provides an independent check  
on \eqref{psiholR}, which was arrived
at in~\cite{Schwarz:2006br} by a rather different line of reasoning from 
\cite{Gunaydin:2006bz}. In particular, the fact that the 
power of $\lambda$ in \eqref{finan} turns out to be opposite to the power
of $p^0$ in \eqref{eguess} is rather non-trivial.

We note that the second factor in \eqref{finan} contributes to genus one 
1-point functions, unless $\alpha=-(n_v+3)/6$. Although such
contributions are perfectly admissible, it is worth noting that the
special value of $\alpha$ where they disappear is also the one where
\eqref{eguess} is invariant under Fourier transform with respect to
all $p^I$ \cite{Pioline:2005vi}.

\end{document}